\documentclass[prd,aps,showpacs,nofootinbib,preprint,eqsecnum]{revtex4}
\usepackage{graphicx,color,amsmath,amsxtra}
\usepackage{epsf}
\usepackage{amssymb}
\usepackage{enumerate}
\usepackage{hhline}
\usepackage{array}
\usepackage{tabularx}

\newcommand{\bear}{\begin{array}}  \newcommand{\eear}{\end{array}}
\newcommand{\bea}{\begin{eqnarray}}  \newcommand{\eea}{\end{eqnarray}}
\newcommand{\beq}{\begin{equation}}  \newcommand{\eeq}{\end{equation}}
\newcommand{\bef}{\begin{figure}}  \newcommand{\eef}{\end{figure}}
\newcommand{\bec}{\begin{center}}  \newcommand{\eec}{\end{center}}

\newcommand{\Eqn}[1]{&\hspace{-0.2em}#1\hspace{-0.2em}&}

\def\Vec#1{\mbox{\boldmath $#1$}}

\def\be{\begin{equation}}
\def\ee{\end{equation}}
\def\bea{\begin{eqnarray}}
\def\eea{\end{eqnarray}}
\def\beq{\begin{eqnarray}}
\def\eeq{\end{eqnarray}}

\def\be{\begin{equation}}
\def\ee{\end{equation}}
\def\bea{\begin{eqnarray}}
\def\eea{\end{eqnarray}}
\def\beq{\begin{eqnarray}}
\def\eeq{\end{eqnarray}}

\begin{document}

\title{Cosmological evolution in exponential gravity
}

\author{Kazuharu Bamba\footnote{E-mail address: 
bamba@phys.nthu.edu.tw}, 
Chao-Qiang Geng\footnote{E-mail address: geng@phys.nthu.edu.tw} 
and 
Chung-Chi Lee\footnote{E-mail address: g9522545@oz.nthu.edu.tw} 
}
\affiliation{
Department of Physics, National Tsing Hua University, Hsinchu, Taiwan 300 
}


\begin{abstract}

We explore the cosmological evolution in the exponential gravity 
$f(R)=R +c_1 \left(1-e^{- c_2 R} \right)$ ($c_{1, 2} = \mathrm{constant}$). 
We summarize various viability conditions and explicitly demonstrate that 
the late-time cosmic acceleration following the matter-dominated stage 
can be realized. 
We also study the equation of state for dark energy and 
confirm that the crossing of the phantom divide from the phantom phase 
to the non-phantom (quintessence) one can occur. 
Furthermore, we illustrate that the cosmological horizon entropy globally 
increases with time. 

\end{abstract}


\maketitle

\section{Introduction}

There exist two representative approaches to account for the current 
accelerated expansion~\cite{WMAP, Komatsu:2010fb, SN1} of the 
universe~\cite{Copeland:2006wr, Nojiri:2006ri, Sotiriou:2008rp, 
DeFelice:2010aj, Review-Tsujikawa, Linder:2010qn}. 
One is to introduce ``dark energy'' in the framework of general relativity. 
The other is to consider a modified gravitational theory, such as 
$f(R)$ gravity. 

Several viable theories of $f(R)$ gravity have been constructed; 
e.g., power-law~\cite{Amendola:2006we, Li:2007xn}, 
Nojiri-Odintsov~\cite{Nojiri-Odintsov}, 
Hu-Sawicki~\cite{Hu:2007nk}, 
Starobinsky's~\cite{Starobinsky:2007hu}, 
Appleby-Battye~\cite{Appleby:2007vb}, 
and Tsujikawa's~\cite{Tsujikawa:2007xu} models 
(for more detailed references, see a recent 
review on $f(R)$ gravity~\cite{DeFelice:2010aj}). 
It is known that these models can satisfy the following conditions for 
the viability: 
(i) positivity of the effective gravitational coupling, 
(ii) stability of cosmological perturbations~\cite{Nojiri:2003ft, 
Dolgov:2003px, Faraoni:2006sy, Song:2006ej}, 
(iii) asymptotic behavior to the standard $\Lambda$-Cold-Dark-Matter 
($\Lambda\mathrm{CDM}$) model in the large curvature regime, 
(iv) stability of the late-time de Sitter point~\cite{Amendola:2006we, 
Muller:1987hp, Faraoni:2005vk}, 
(v) constraints from the equivalence principle, 
and 
(vi) solar-system constraints~\cite{Solar-System-Constraints}.

Recently, an interesting model of 
$f(R)=R +c_1 \left(1-e^{- c_2 R} \right)$, 
called ``exponential gravity'', has been proposed in 
Refs.~\cite{Exponential-type-f(R)-gravity, Cognola:2007zu, Linder:2009jz} 
with $c_{1, 2}$ being constants. 
The important feature of the exponential gravity is that it has only one 
more parameter than the $\Lambda\mathrm{CDM}$ model. 
The constraints from the violation of the equivalence 
principle~\cite{Tsujikawa:2009ku} and 
cosmological observations~\cite{Ali:2010zx} on the exponential gravity 
have been examined. 
The exponential gravity in the framework of $f(R)$ gravity has been extended 
to a gravitational theory in terms of the torsion scalar~\cite{Linder:2010py} 
(for a related work on torsion gravity, see~\cite{Bengochea:2008gz}). 
We note that the cosmological dynamics in the gravitational theory 
consisting only of the exponential term without the Einstein-Hilbert one 
has also been studied in Ref.~\cite{Abdelwahab:2007jp}. 

In this paper, we explicitly investigate the cosmological evolution 
in the exponential gravity model given by 
Cognola {\it et al.}~\cite{Cognola:2007zu} 
and 
Linder~\cite{Linder:2009jz} 
in more detail by using the analysis method in Ref.~\cite{Hu:2007nk}. 
We also check the above six viability conditions for the model. 
In particular, 
we demonstrate that after the matter-dominated stage, 
the current accelerated expansion of the universe and 
the crossing of the phantom divide from the phantom phase to 
the non-phantom (quintessence) one can be realized. 
It is interesting to note that 
the crossing of the phantom divide is implied by the cosmological 
observational data~\cite{observational status}, 
while the exponential gravity is a ghost free theory. 
In addition, we illustrate that the cosmological horizon entropy globally 
increases with time. 
We use units of 
$k_\mathrm{B} = c = \hbar = 1$ and 
denote the gravitational constant $8 \pi G$ by 
${\kappa}^2 \equiv 8\pi/{M_{\mathrm{Pl}}}^2$ 
with the Planck mass of 
$M_{\mathrm{Pl}} = G^{-1/2} = 1.2 \times 10^{19}$\,\,GeV.

The paper is organized as follows. 
In Sec.\ II, we review the model of the exponential gravity in 
Refs.~\cite{Cognola:2007zu, Linder:2009jz} 
and summarize its viability conditions. 
In Sec.\ III, we explore the cosmological evolution of the model. 
We examine the horizon entropy in Sec.\ IV. 
Finally, conclusions are given in Sec.\ V.

\section{Exponential gravity}

\subsection{The model}

The action of $f(R)$ gravity with matter is given by
\begin{eqnarray}
I = \int d^4 x \sqrt{-g} \frac{f(R)}{2\kappa^2} + I_{\mathrm{matter}} 
(g_{\mu\nu}, \Upsilon_{\mathrm{matter}})\,,
\label{eq:2.1}
\end{eqnarray}
where $g$ is the determinant of the metric tensor $g_{\mu\nu}$, 
$I_{\mathrm{matter}}$ is the action of matter which is assumed to be minimally 
coupled to gravity, i.e., the action $I$ is written in the Jordan frame, 
and $\Upsilon_{\mathrm{matter}}$ denotes matter fields. 
Here, we use the standard metric formalism. 

Taking the variation of the action in Eq.~(\ref{eq:2.1}) with respect to 
$g_{\mu\nu}$, one obtains~\cite{Sotiriou:2008rp} 
\begin{equation}
F G_{\mu\nu} 
= 
\kappa^2 T^{(\mathrm{matter})}_{\mu \nu} 
-\frac{1}{2} g_{\mu \nu} \left( FR - f \right)
+ \nabla_{\mu}\nabla_{\nu}F -g_{\mu \nu} \Box F\,,
\label{eq:2.2}
\end{equation}
where 
$G_{\mu\nu}=R_{\mu\nu}-\left(1/2\right)g_{\mu\nu}R$ is 
the Einstein tensor, 
$F(R) \equiv d f(R)/dR$, 
${\nabla}_{\mu}$ is the covariant derivative operator associated with 
$g_{\mu \nu}$, 
$\Box \equiv g^{\mu \nu} {\nabla}_{\mu} {\nabla}_{\nu}$
is the covariant d'Alembertian for a scalar field, 
and 
$T^{(\mathrm{matter})}_{\mu \nu}$ 
is the contribution to the energy-momentum tensor from all 
perfect fluids of generic matter. 

In this paper, 
we concentrate on the exponential gravity 
in Refs.~\cite{Cognola:2007zu, Linder:2009jz}, 
given by 
\begin{eqnarray}
f(R) = R -\beta R_{\mathrm{s}}\left(1-e^{-R/R_{\mathrm{s}}} \right) \,,
\label{eq:2.12}
\end{eqnarray}
where $c_1 = -\beta R_{\mathrm{s}}$ and $c_2 = R_{\mathrm{s}}^{-1}$. 
Note that $R_{\mathrm{s}}$ corresponds to the characteristic 
curvature modification scale.

\subsection{Viability conditions on exponential gravity}

For the model of the exponential gravity in Eq.~(\ref{eq:2.12}), 
it is straightforward to show that the conditions for the 
viability can be satisfied, which are summarized as follows: 

\noindent
{\bf (i) Positivity of the effective gravitational coupling}
\vspace{1mm}

When $\beta<e^{R/R_{\mathrm{s}}}$, $F(R) = 1-\beta e^{-R/R_{\mathrm{s}}} > 0$. 
This is required for the positivity of the effective gravitational coupling 
$G_{\mathrm{eff}} \equiv G/F(R) >0$ to avoid anti-gravity. 
In the sense of the quantum theory, the graviton is not a ghost. 

\noindent
{\bf (ii) Stability of cosmological perturbations}
\vspace{1mm}

When $\beta>0$ and $R_{\mathrm{s}}>0$, 
$f^{\prime \prime}(R) = F^{\prime}(R) = 
\left(\beta/R_{\mathrm{s}}\right)e^{-R/R_{\mathrm{s}}} >0$, 
where the prime denotes differentiation with respect to $R$. 
This is required for the stability of cosmological 
perturbations~\cite{Dolgov:2003px, Faraoni:2006sy, Song:2006ej}. 
In the sense of the quantum theory, the scalaron, which is a new scalar degree 
of freedom in $f(R)$ gravity, is not a tachyon~\cite{Starobinsky:2007hu}. 

\noindent
{\bf (iii) Asymptotic behavior to the $\Lambda\mathrm{CDM}$ model 
in the large curvature regime}
\vspace{1mm}

Since $f(R) - R \to -\beta R_{\mathrm{s}} = \mathrm{constant}$ for 
$R/R_{\mathrm{s}} \gg 1$, 
this model is reduced to the $\Lambda\mathrm{CDM}$ model in the large 
curvature regime $R/R_{\mathrm{s}} \gg 1$. Such a behavior is necessary for 
the presence of the matter-dominated stage. 

\noindent
{\bf (iv) Stability of the late-time de Sitter point}
\vspace{1mm}

When $\beta>1$, 
$0<m(R=R_{\mathrm{d}})<1$~\cite{Tsujikawa:2009ku}, where 
$m \equiv R f^{\prime \prime}(R)/f^{\prime}(R) = RF^{\prime}(R)/F(R)$ and 
$R_{\mathrm{d}} = 2f(R_{\mathrm{d}})/F(R_{\mathrm{d}})$ 
is the value of the scalar curvature at the de Sitter point. 
This condition is required for the stability of the late-time de Sitter 
point~\cite{Amendola:2006we, Muller:1987hp, Faraoni:2005vk}. 
The quantity $m$ characterizes the deviation from 
the $\Lambda\mathrm{CDM}$ model because $m=0$ 
for the $\Lambda\mathrm{CDM}$ model. 
In the exponential gravity, 
by using $m(R)= \beta \left(R/R_{\mathrm{s}}\right)e^{-R/R_{\mathrm{s}}}/
\left(1-\beta e^{-R/R_{\mathrm{s}}}\right)$ 
and 
$\beta = \left(R_{\mathrm{d}}/R_{\mathrm{s}}\right)/\left[
2-\left(2+R_{\mathrm{d}}/R_{\mathrm{s}}\right)
e^{-R_{\mathrm{d}}/R_{\mathrm{s}}}\right]$, 
one finds that $m(R_{\mathrm{d}})<1$ for $R_{\mathrm{d}}/R_{\mathrm{s}}>0$. 
Hence, $m(R_{\mathrm{d}})<1$ for $\beta>1$. 

\noindent
{\bf (v) Constraints from the violation of the equivalence principle}
\vspace{1mm}

It is known that $f(R)$ gravity can satisfy local gravity constraints from the 
violation of the equivalence principle under the chameleon 
mechanism~\cite{Mota:2003tc, Chameleon mechanism}. 
By making the following conformal 
transformation~\cite{Conformal Transformation}: 
$
g_{\mu \nu} \, \rightarrow \, 
\tilde{g}_{\mu \nu} = \Xi^2 g_{\mu \nu}
$, 
the action of $f(R)$ gravity in Eq.~(\ref{eq:2.1}) can be 
rewritten in the Einstein frame, 
where
$
\Xi^2 \equiv F = e^{\sqrt{2/3} \kappa \phi}
$ 
with the scalar field $\phi$. 
In what follows, a tilde represents the quantity in the Einstein frame. 
We consider a spherically symmetric body with radius $\tilde{r}_{\mathrm{c}}$ 
in the Minkowski space-time. 
Here, $\tilde{r}$ is the distance of the center of the body, and 
$\rho^{*}=e^{-\sqrt{3/2}\kappa\phi}\rho$ is a conserved matter 
density in the Einstein frame with $\rho$ the energy density of 
matter in the Jordan frame. 
We assume that a spherically symmetric body has constant densities of 
$\rho^{*}=\rho_{\mathrm{in}}$ 
and $\rho_{\mathrm{out}} (\ll \rho_{\mathrm{in}})$ 
inside ($\tilde{r}<\tilde{r}_{\mathrm{c}}$) 
and outside ($\tilde{r}>\tilde{r}_{\mathrm{c}}$), respectively. 
In this case, the effective potential has two minima at the field values 
$\phi_{\mathrm{in}}$ and $\phi_{\mathrm{out}}$ satisfying 
the conditions $dV_{\mathrm{eff}}(\phi_{\mathrm{in}})/d\phi = 0$ and 
$dV_{\mathrm{eff}}(\phi_{\mathrm{out}})/d\phi = 0$ 
with a heavier mass squared 
$m_{\mathrm{in}}^2 \equiv d^2V_{\mathrm{eff}}(\phi_{\mathrm{in}})/d\phi^2$
and a lighter mass squared 
$m_{\mathrm{out}}^2 \equiv d^2V_{\mathrm{eff}}(\phi_{\mathrm{out}})/d\phi^2$, 
respectively. 
The thin-shell parameter is defined as 
$
\epsilon_{\mathrm{th}} \equiv - \kappa
\left(\phi_{\mathrm{out}}-\phi_{\mathrm{in}}\right)/
\left(\sqrt{6}\Phi_{\mathrm{c}}\right)
$~\cite{Chameleon mechanism}, 
where 
$\Phi_{\mathrm{c}} = GM_{\mathrm{c}}/\tilde{r}_{\mathrm{c}}$ 
is the gravitational potential at the surface of the body 
and 
$M_{\mathrm{c}}=\left(4\pi/3\right)\tilde{r}_{\mathrm{c}}^3
\rho_{\mathrm{in}}$. 

The tightest experimental bound on $\epsilon_{\mathrm{th}}$ obtained from 
the violation of the equivalence principle for the accelerations of the Earth 
and the moon toward the Sun is given by 
$
\epsilon_{\mathrm{th},\,\oplus} < 2.2 \times 10^{-6}
$~\cite{Capozziello:2007eu, Tsujikawa:2008uc}. 
This is the thin-shell parameter for the Earth. 
By using the value of the gravitational potential for the Earth 
$\Phi_{\mathrm{c}\,,\oplus} = 7.0 \times 10^{-10}$ and 
$|\phi_{\mathrm{out},\,\oplus}| \gg |\phi_{\mathrm{in},\,\oplus}|$, 
the condition on $\epsilon_{\mathrm{th},\,\oplus}$ is reduced to 
$
|\kappa \phi_{\mathrm{out},\,\oplus}| < 3.7 \times 10^{-15}
$~\cite{Tamaki:2008mf}. 
The field value $\phi_{\mathrm{out},\,\oplus}$ can be found by solving 
$dV_{\mathrm{eff}}(\phi_{\mathrm{out}})/d\phi = 0$ with 
$\rho^{*}=\rho_{\mathrm{out}}$, which gives $R \simeq \kappa^2 
\rho_{\mathrm{out}}$. 

For the exponential gravity, 
$\kappa \phi_{\mathrm{out}} \simeq -\sqrt{3/2} \beta 
e^{-\kappa^2 \rho_{\mathrm{out}}/R_{\mathrm{s}}}$~\cite{Tsujikawa:2009ku} 
and $\beta R_{\mathrm{s}}/R_0 \approx \Omega_{\mathrm{m}}^{(0)}$, 
where $R_0 \approx 12H_0^2$ is the current scalar curvature, 
$H_0$ is the current Hubble parameter, 
$\Omega_{\mathrm{m}}^{(0)} \equiv 
\rho_{\mathrm{m}}^{(0)}/\rho_{\mathrm{crit}}^{(0)}$ 
is the current density parameter of non-relativistic matter 
(cold dark matter and baryon), 
$\rho_{\mathrm{m}}^{(0)}$ is the energy density of non-relativistic matter 
at the present time, 
and $\rho_{\mathrm{crit}}^{(0)} = 3H_0^2/\kappa^2$ 
is the critical density. 
As a consequence, by using $\rho_{\mathrm{crit}}^{(0)} \simeq 10^{-29}\, 
\mathrm{g}/\mathrm{cm}^3$ and 
the homogeneous baryon/dark matter density 
$\rho_{\mathrm{out}} \simeq 10^{-24}\, \mathrm{g}/\mathrm{cm}^3$, 
we find 
$\kappa \phi_{\mathrm{out}} 
\approx -\beta \exp\left(-10^{5}\beta \right)$~\cite{Tsujikawa:2009ku}. 
When $\beta>1$, which is the stability condition for 
the late-time de Sitter point in the exponential gravity, the above constraint 
on $|\kappa \phi_{\mathrm{out},\,\oplus}|$ is satisfied very well. 
For example, if $\beta=1.1$, 
$|\kappa \phi_{\mathrm{out}}| = O(10^{-50000})$. 
In what follows, the superscript $(0)$ denotes the present value. 

\noindent
{\bf (vi) Solar-system constraints}
\vspace{1mm}

The bound on the thin-shell parameter 
coming from the solar-system constraint 
$\epsilon_{\mathrm{th},\,\odot} < 2.3 \times 10^{-5}$~\cite{DeFelice:2010aj} 
is weaker than that from the violation of the equivalence principle 
$\epsilon_{\mathrm{th},\,\oplus} < 2.2 \times 10^{-6}$ shown above.

\section{Cosmological evolution}

We assume the flat Friedmann-Lema\^{i}tre-Robertson-Walker (FLRW) 
space-time with the metric, 
\begin{eqnarray}
{ds}^2 = -{dt}^2 + a^2(t)d{\Vec{x}}^2\,,
\label{eq:3.1}
\end{eqnarray}
where $a(t)$ is the scale factor. 
{}From Eq.~(\ref{eq:2.2}), 
we obtain the following gravitational field equations: 
\begin{eqnarray} 
3FH^2 
\Eqn{=}
\kappa^2 \rho_{\mathrm{M}} +\frac{1}{2} \left( FR - f \right) 
-3H\dot{F}\,,
\label{eq:3.2} \\ 
-2F\dot{H}  
\Eqn{=}
\kappa^2 \left( \rho_{\mathrm{M}} + P_{\mathrm{M}} \right)
+\ddot{F}-H\dot{F}\,,
\label{eq:3.3}
\end{eqnarray} 
where $H=\dot{a}/a$ is the Hubble parameter, 
the dot denotes the time derivative of $\partial/\partial t$, and 
$\rho_{\mathrm{M}}$ and $P_{\mathrm{M}}$ are 
the energy density and pressure of all perfect fluids of generic matter, 
respectively. 

Equation (\ref{eq:3.2}) can be rewritten to 
\begin{equation}
H^2 -\left(F-1\right)\left(H \frac{dH}{d \ln a} + H^2\right) 
+\frac{1}{6}\left(f-R\right) + H^2 F^{\prime} \frac{dR}{d \ln a} 
= \frac{\kappa^2 \rho_{\mathrm{M}}}{3}\,,
\label{eq:3.4}
\end{equation}
while the scalar curvature $R$ is expressed as 
\begin{equation}
R=12H^2+6H\frac{dH}{d \ln a}\,.
\label{eq:3.5}
\end{equation}
To solve Eqs.~(\ref{eq:3.4}) and (\ref{eq:3.5}), we introduce the following 
variables~\cite{Hu:2007nk}: 
\begin{eqnarray} 
y_H \Eqn{\equiv} \frac{\rho_{\mathrm{DE}}}{\rho_{\mathrm{m}}^{(0)}}
= \frac{H^2}{\bar{m}^2} -a^{-3} -\chi a^{-4}\,,
\label{eq:3.6} \\ 
y_R \Eqn{=} \frac{R}{\bar{m}^2} -3a^{-3}\,,
\label{eq:3.7}
\end{eqnarray} 
with
\begin{eqnarray} 
\bar{m}^2 \Eqn{\equiv} \frac{\kappa^2 \rho_{\mathrm{m}}^{(0)}}{3}\,,
\label{eq:3.8} \\ 
\chi \Eqn{\equiv} \frac{\rho_{\mathrm{r}}^{(0)}}{\rho_{\mathrm{m}}^{(0)}}
\simeq 3.1 \times 10^{-4}\,, 
\label{eq:3.9}
\end{eqnarray} 
where $\rho_{\mathrm{DE}}$ is the energy density of dark energy 
and $\rho_{\mathrm{r}}^{(0)}$ is the energy density of radiation 
at the present time. 
In our analysis, the contribution from radiation is also taken into 
consideration. 
Equations (\ref{eq:3.4}) and (\ref{eq:3.5}) are reduced to 
a coupled set of ordinary differential equations 
\begin{eqnarray}
\frac{dy_H}{d \ln a} 
\Eqn{=} 
\frac{y_R}{3}-4y_H\,,
\label{eq:3.10} \\ 
\frac{dy_R}{d \ln a} 
\Eqn{=} 9a^{-3} -\frac{1}{y_H+a^{-3}+\chi a^{-4}}\frac{1}{\bar{m}^2 F^{\prime}}
\nonumber \\ 
&&
{}
\times \left[ y_H -\left(F-1\right) \left(\frac{1}{6}y_R-y_H-\frac{1}{2}a^{-3}
-\chi a^{-4}\right) + \frac{1}{6}\frac{f-R}{\bar{m}^2} \right]\,.
\label{eq:3.11}
\end{eqnarray} 

The equation of state for dark energy $w_{\mathrm{DE}} \equiv 
P_{\mathrm{DE}}/\rho_{\mathrm{DE}}$ 
is given by 
\begin{equation}
w_{\mathrm{DE}} 
= -1 -\frac{1}{3}\frac{1}{y_H}\frac{dy_H}{d \ln a}\,,
\label{eq:3.12}
\end{equation}
derived by the continuity equation 
\begin{equation}
\dot{\rho}_{\mathrm{DE}} 
+ 3H\left(1+w_{\mathrm{DE}}\right)\rho_{\mathrm{DE}} = 0\,.
\label{eq:3.13}
\end{equation}
%
On the other hand, the effective equation of state 
$w_{\mathrm{eff}}$ is defined as 
\begin{equation}
w_{\mathrm{eff}} 
\equiv 
-1 -\frac{2}{3}\frac{\dot{H}}{H^2} 
= \frac{P_{\mathrm{tot}}}{\rho_{\mathrm{tot}}}
\,,
\label{eq:3.A-1}
\end{equation}
where $\rho_{\mathrm{tot}} \equiv \rho_{\mathrm{DE}} + \rho_{\mathrm{m}} + 
\rho_{\mathrm{r}}$ and 
$P_{\mathrm{tot}} \equiv P_{\mathrm{DE}} + P_{\mathrm{m}} + P_{\mathrm{r}}$ 
are the total energy density and pressure of the universe, 
respectively. Here, $P_{\mathrm{DE}}$, $P_{\mathrm{m}} \, (= 0)$ and 
$P_{\mathrm{r}}$ are the pressure of dark energy, non-relativistic matter 
and radiation, respectively. 

Combining Eqs.~(\ref{eq:3.10}) and (\ref{eq:3.11}), we obtain 
\begin{equation}
\frac{d^2y_H}{d \left(\ln a\right)^2} 
+J_1 \frac{dy_H}{d \ln a} +J_2 y_H + J_3=0\,,
\label{eq:3.14}
\end{equation} 
where 
\begin{eqnarray}
J_1 \Eqn{=} 
4+\frac{1}{y_H+a^{-3}+\chi a^{-4}}
\frac{1-F}{6\bar{m}^2 F^{\prime}}\,,
\label{eq:3.15} \\
J_2 \Eqn{=}
\frac{1}{y_H+a^{-3}+\chi a^{-4}}
\frac{2-F}{3\bar{m}^2 F^{\prime}}\,,
\label{eq:3.16} \\
J_3 \Eqn{=} 
-3a^{-3}-\frac{ \left(1-F\right) \left(a^{-3}+2\chi a^{-4}\right) 
+\left(R-f\right)/\left(3\bar{m}^2\right)}{y_H+a^{-3}+\chi a^{-4}}
\frac{1}{6\bar{m}^2 F^{\prime}}\,.
\label{eq:3.17}
\end{eqnarray} 

In Figs.~1, 2 and 3, 
we depict the cosmological evolutions of the density parameters of dark energy 
$\Omega_{\mathrm{DE}} \equiv \rho_{\mathrm{DE}}/\rho_{\mathrm{crit}}^{(0)}$, 
non-relativistic matter 
$\Omega_{\mathrm{m}} \equiv 
\rho_{\mathrm{m}}/\rho_{\mathrm{crit}}^{(0)}$ 
and radiation $\Omega_{\mathrm{r}} \equiv 
\rho_{\mathrm{r}}/\rho_{\mathrm{crit}}^{(0)}$ 
as functions of the redshift $z \equiv 1/a -1$ 
for $\beta=1.1$, $\beta=1.8$ and $\beta=2.5$, 
respectively. 
In the high $z$ regime ($z \gtrsim 3.0$), the universe is at the 
matter-dominated stage ($\Omega_{\mathrm{m}} > \Omega_{\mathrm{DE}}$, 
$\Omega_{\mathrm{m}} \gg \Omega_{\mathrm{r}}$). 
As $z$ decreases, the dark energy becomes dominant over matter for 
$z<z_{\mathrm{DE}}$, 
where $z_{\mathrm{DE}}$ is the crossover point in which 
$\Omega_{\mathrm{DE}} = \Omega_{\mathrm{m}}$. 
Explicitly, we have 
$z_{\mathrm{DE}} = 0.55$, $0.47$ and $0.45$ for 
$\beta=1.1$, $1.8$ and $2.5$, respectively. 
The values of $z_{\mathrm{DE}}$ become smaller for the larger values of 
$\beta$. 
At the present time ($z=0$), 
$(\Omega_{\mathrm{DE}}^{(0)}, \Omega_{\mathrm{m}}^{(0)}, 
\Omega_{\mathrm{r}}^{(0)}) = 
(0.77, 0.23, 7.0 \times 10^{-5})$, $(0.76, 0.24, 7.3 \times 10^{-5})$ and 
$(0.75, 0.25, 7.3 \times 10^{-5})$ for 
$\beta=1.1$, $1.8$ and $2.5$, respectively. 
In Fig.~4, we also show the cosmological evolution of $\Omega_{\mathrm{r}}$ 
for $\beta=1.8$. 
The qualitative behaviors of $\Omega_{\mathrm{r}}$ for $\beta=1.1$ and $2.5$ 
are similar to that for $\beta=1.8$. 
Thus, the current accelerated expansion of the universe following 
the matter-dominated stage can be realized in the exponential gravity.

\begin{figure}[tbp]
\begin{center}
\resizebox{!}{13cm}{
   \includegraphics{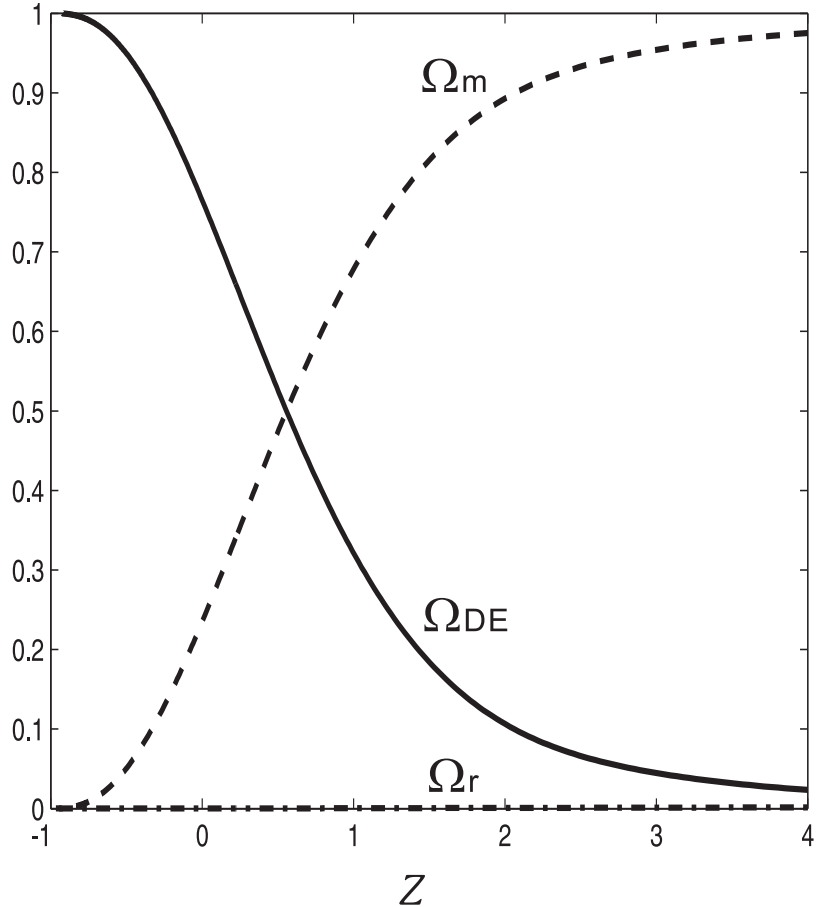}
                  }
\caption{Cosmological evolutions of 
$\Omega_{\mathrm{DE}}$ (solid line), $\Omega_{\mathrm{m}}$ (dashed line) and 
$\Omega_{\mathrm{r}}$ (dashed and single-dotted line) 
as functions of the redshift $z$ 
for $\beta=1.1$. 
}
\end{center}
\label{fg:1}
\end{figure}

\begin{figure}[htbp]
\begin{center}
\resizebox{!}{13cm}{
   \includegraphics{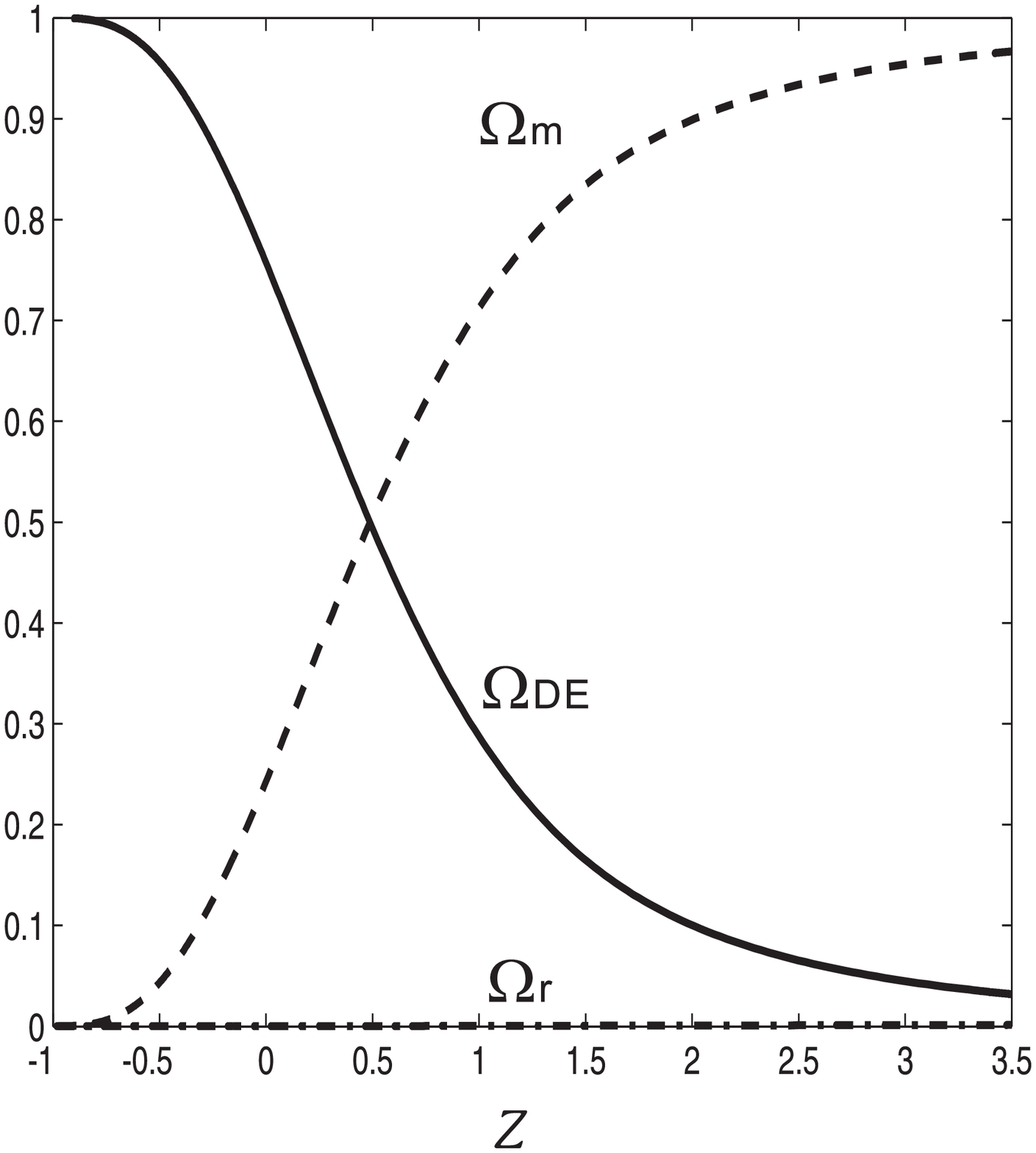}
                  }
\caption{Legend is the same as Fig.~1 but for $\beta=1.8$. 
}
\end{center}
\label{fg:2}
\end{figure}

\begin{figure}[htbp]
\begin{center}
\resizebox{!}{13cm}{
   \includegraphics{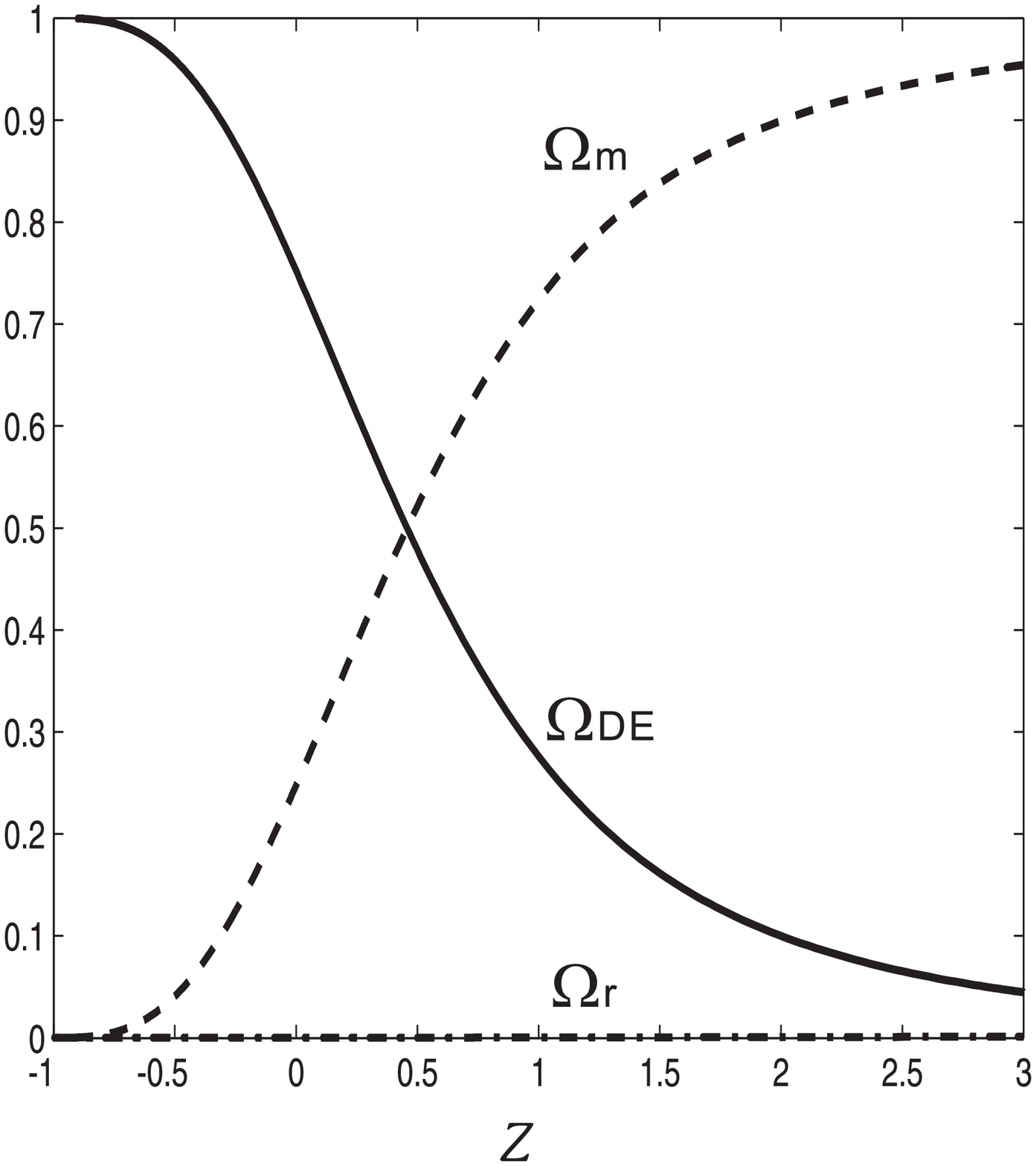}
                  }
\caption{Legend is the same as Fig.~1 but for $\beta=2.5$. 
}
\end{center}
\label{fg:3}
\end{figure}

\begin{figure}[htbp]
\begin{center}
\resizebox{!}{13cm}{
   \includegraphics{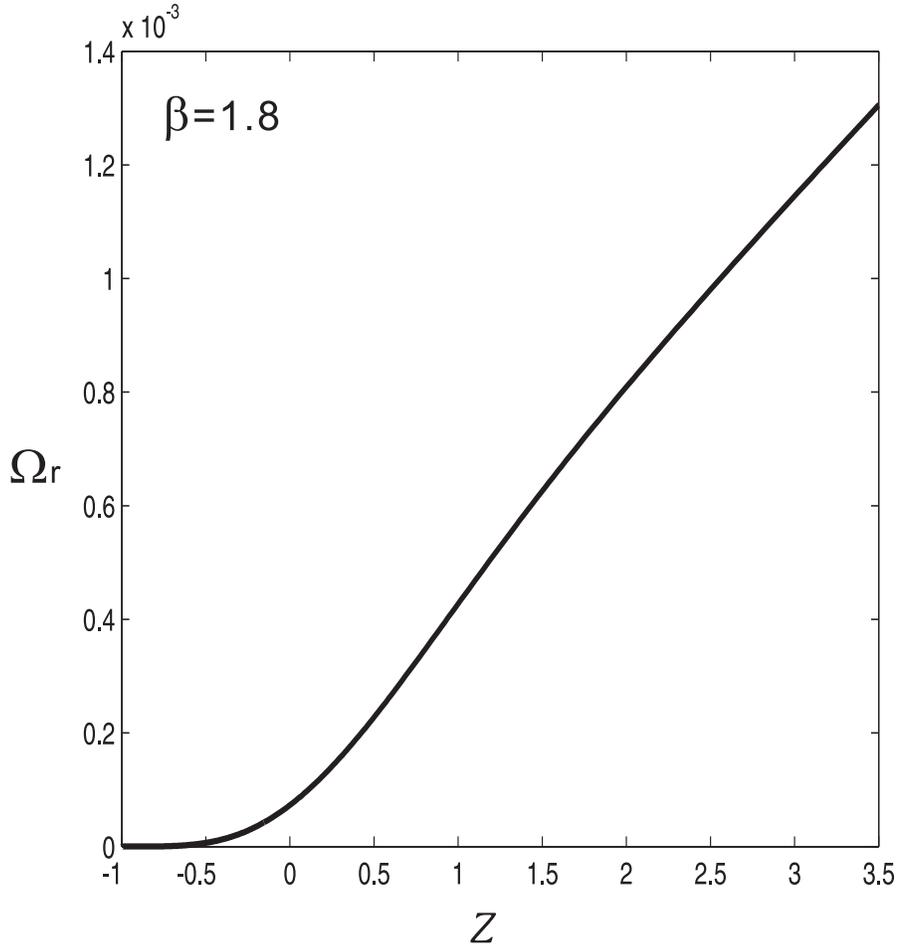}
                  }
\caption{Cosmological evolution of $\Omega_{\mathrm{r}}$ (solid line) 
as a function of the redshift $z$ for $\beta=1.8$. 
}
\end{center}
\label{fg:4}
\end{figure}

\begin{figure}[htbp]
\begin{center}
\resizebox{!}{13cm}{
   \includegraphics{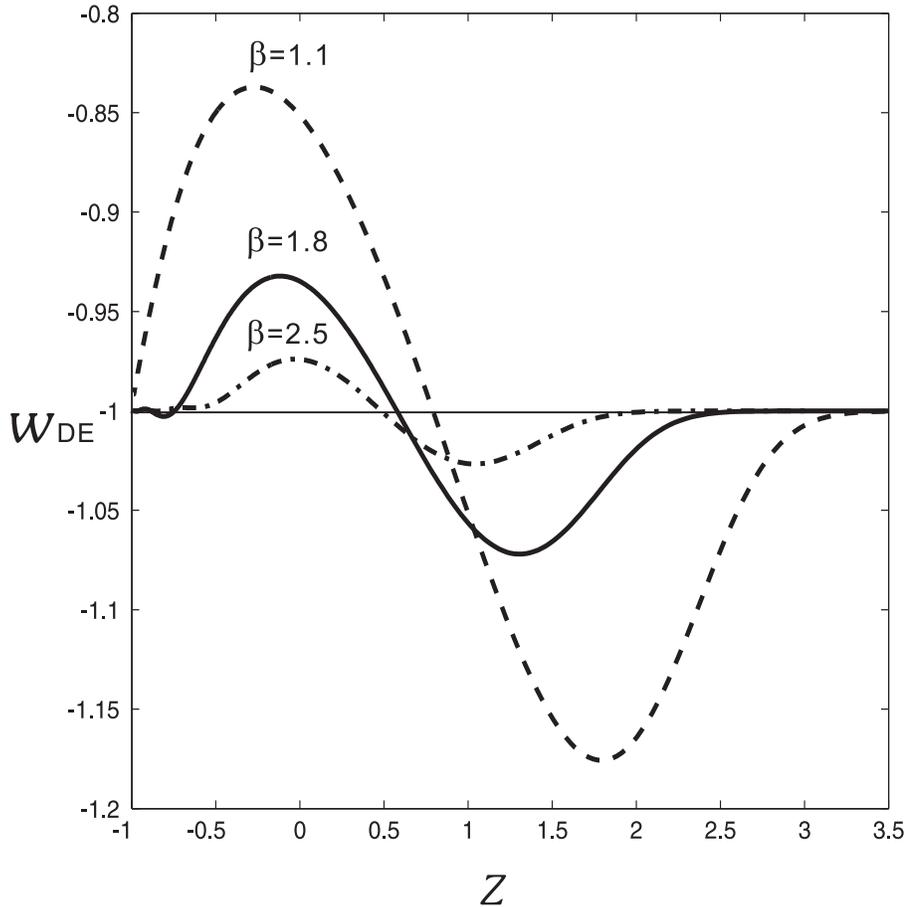}
                  }
\caption{Cosmological evolution of $w_{\mathrm{DE}}$ in Eq.~(\ref{eq:3.12}) 
as a function of the redshift $z$ 
for $\beta=1.1$ (dashed line), $\beta=1.8$ (thick solid line) 
and $\beta=2.5$ (dashed and single-dotted line), 
where the thin solid line shows $w_{\mathrm{DE}}=-1$ (cosmological constant). 
}
\end{center}
\label{fg:5}
\end{figure}

\begin{figure}[htbp]
\begin{center}
\resizebox{!}{13cm}{
   \includegraphics{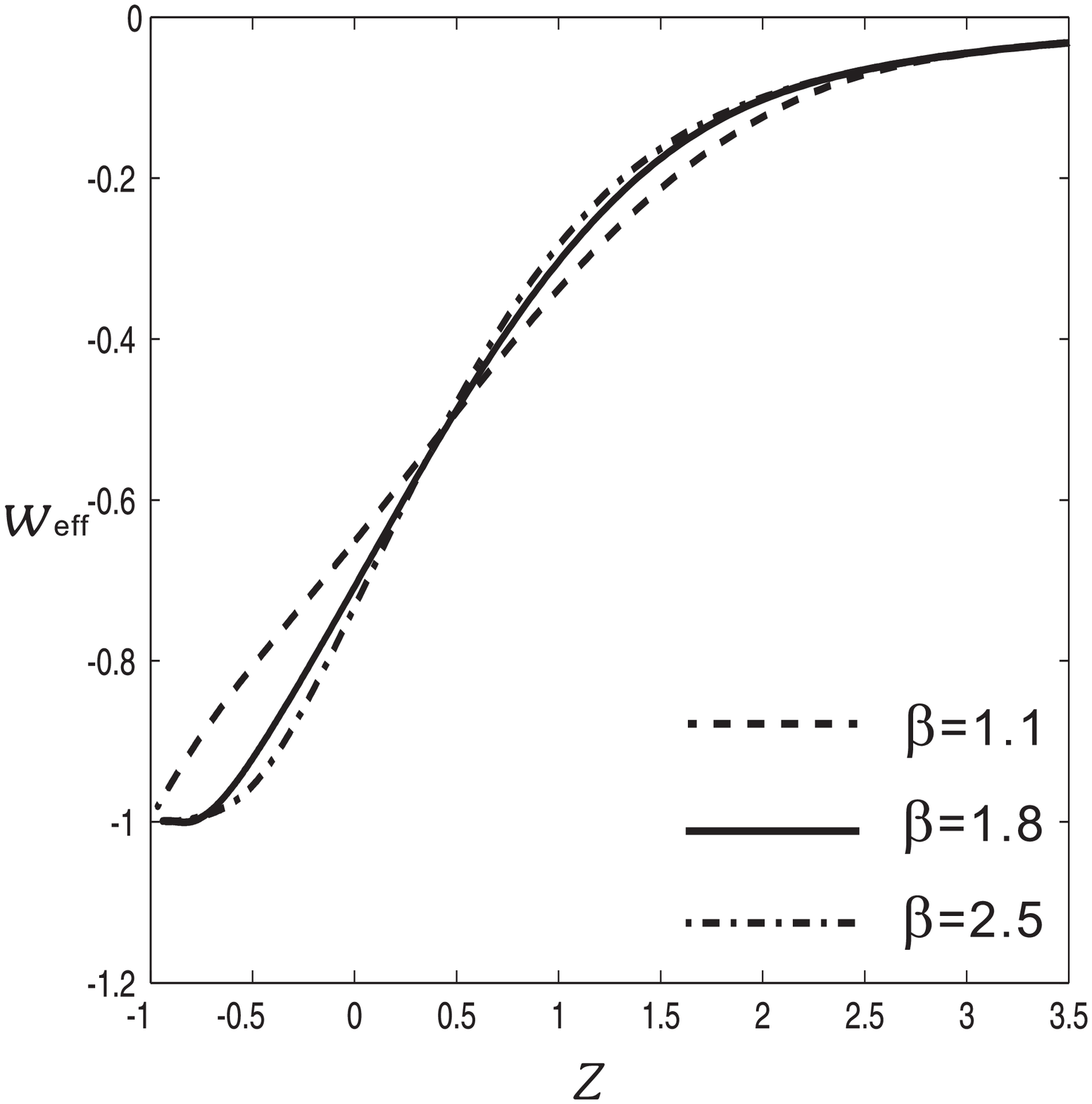}
                  }
\caption{Cosmological evolution of $w_{\mathrm{eff}}$ in Eq.~(\ref{eq:3.A-1}) 
as a function of the redshift $z$ 
for $\beta=1.1$ (dashed line), $\beta=1.8$ (thick solid line) 
and $\beta=2.5$ (dashed and single-dotted line). 
}
\end{center}
\label{fg:6}
\end{figure}

We note that in solving Eq.~(\ref{eq:3.17}) numerically, 
we have taken the initial conditions 
at $z=z_{\mathrm{i}}$ 
as $y_H \, (z=z_{\mathrm{i}}) = 3.0$ and 
$dy_H/d \ln a \, (z=z_{\mathrm{i}}) = 0$, 
where $z_{\mathrm{i}}=4.0$, $3.5$ and $3.0$ for $\beta=1.1$, $\beta=1.8$ 
and $\beta=2.5$, respectively. 
The values of $z_{\mathrm{i}}$ have been chosen so that 
$RF^{\prime}(z=z_{\mathrm{i}}) \sim 10^{-13}$, i.e., 
the exponential gravity at $z=z_{\mathrm{i}}$ can be very close to 
the $\Lambda\mathrm{CDM}$ model, in which $RF^{\prime} =0$. 
Since $R/R_{\mathrm{s}} \gg 1$ in the high $z$ regime 
($z \simeq z_{\mathrm{i}}$), 
$\beta R_{\mathrm{s}}/\bar{m}^2 \simeq 6y_H$. 
Consequently, the value of the combination $\beta R_{\mathrm{s}}$ is set as 
$\beta R_{\mathrm{s}} \simeq 18H_0^2 \Omega_{\mathrm{m}}^{(0)}$. 
Therefore, we have only one free parameter $\beta$ in the exponential gravity 
in Eq.~(\ref{eq:2.12}). 
Furthermore, 
from Eq.~(\ref{eq:3.10}) we see that $y_R = 12 y_H$ at $z=z_{\mathrm{i}}$ 
and it follows from Eq.~(\ref{eq:3.12}) that 
$w_{\mathrm{DE}} = -1$ at $z=z_{\mathrm{i}}$. 
All numerical calculations have been executed for 
$\Omega_{\mathrm{m}}^{(0)} = 0.26$~\cite{Komatsu:2010fb}. 

The cosmological evolution of the equation of state for dark energy 
$w_{\mathrm{DE}}$ in Eq.~(\ref{eq:3.12}) is shown in Fig.~5. 
{}From the figure, we see that $w_{\mathrm{DE}}$ starts at 
the phase of a cosmological constant $w_{\mathrm{DE}} = -1$ 
and evolves from the phantom phase ($w_{\mathrm{DE}} < -1$) to 
the non-phantom (quintessence) phase ($w_{\mathrm{DE}} > -1$). 
The crossing of the phantom divide occurs at $z = z_{\mathrm{cross}}$, 
where $z_{\mathrm{cross}} = 0.78$, $0.57$ and $0.46$ for 
$\beta=1.1$, $1.8$ and $2.5$, respectively. 
The values of $z_{\mathrm{cross}}$ become smaller for the larger values 
of $\beta$. 
Moreover, 
the present values of $w_{\mathrm{DE}}$ 
are $w_{\mathrm{DE}} (z=0) = -0.85$, $-0.93$ and $-0.97$ for 
$\beta=1.1$, $1.8$ and $2.5$, respectively. 
Since $\beta R_{\mathrm{s}}$ is a constant, 
the larger $\beta$ is, the closer the exponential gravity is 
to the $\Lambda\mathrm{CDM}$ model. 
The results on $w_{\mathrm{DE}}$ are qualitatively the same as 
the analysis in Refs.~\cite{Linder:2009jz, Ali:2010zx}. 
Thus, the crossing of the phantom divide from the phantom phase to the 
non-phantom one can be realized in the exponential gravity. 
We remark that the similar behaviors can occur in 
Hu-Sawicki~\cite{Hu:2007nk, Martinelli:2009ek}, 
Appleby-Battye~\cite{Appleby:2009uf}, and 
Starobinsky's~\cite{M-S-Y} models 
as well. 

In Fig.~6, we also illustrate the cosmological evolution of the effective 
equation of state $w_{\mathrm{eff}}$ in Eq.~(\ref{eq:3.A-1}). 
The present values of $w_{\mathrm{eff}}$
are $w_{\mathrm{eff}} (z=0) = -0.65$, $-0.71$ and $-0.74$ for 
$\beta=1.1$, $1.8$ and $2.5$, respectively. 
We remark that $w_{\mathrm{eff}}$ does not cross the line of 
the phantom divide unlike $w_{\mathrm{DE}}$ due to 
the null energy condition 
$\rho_{\mathrm{tot}} + P_{\mathrm{tot}} = 
\rho_{\mathrm{DE}} + \rho_{\mathrm{m}} + \rho_{\mathrm{r}} 
+ P_{\mathrm{DE}} + P_{\mathrm{m}} + P_{\mathrm{r}} \geq 0$. 

Finally, we mention that an $f(R)$ gravity model with realizing a crossing of 
the phantom divide from the non-phantom phase to the phantom one, which is 
the opposite transition from the above one, has been reconstructed in 
Ref.~\cite{BGNO-PC}. 
In addition, 
the behavior of $f(R)$ gravity with realizing multiple crossings of the 
phantom divide~\cite{Bamba:2009kc} and that of $f(R)$ gravity around a 
crossing of the phantom divide by taking into account the presence of cold 
dark matter~\cite{Bamba:2009dk} have also been explored.

\section{Horizon entropy}

In Ref.~\cite{Bamba:2009id}, 
it is shown that it is possible to obtain a picture of equilibrium 
thermodynamics on the apparent horizon in the FLRW background 
for $f(R)$ gravity as well as that of non-equilibrium thermodynamics 
due to a suitable redefinition of an energy momentum tensor of 
the ``dark'' component that respects a local energy conservation. 
For a recent review on the Black hole entropy on scalar-tensor and 
$f(R)$ gravity, see Ref.~\cite{Faraoni:2010yi}. 

In general relativity, the Bekenstein-Hawking horizon entropy is expressed as 
$S=A/\left(4G\right)$, where $A$ is the area of the 
apparent horizon~\cite{Bekenstein:1973ur, Bardeen:1973gs, Hawking:1974sw}. 
The Bekenstein-Hawking entropy 
\begin{equation}
S=\frac{A}{4G}
\label{eq:4.B-1}
\end{equation}
is a global geometric quantity which is proportional to the horizon area $A$ 
with a constant coefficient $1/\left(4G\right)$. 
This quantity is not directly affected by the difference of 
gravitational theories. 
We regard the horizon entropy $S$ in Eq.~(\ref{eq:4.B-1}) as the one 
in the equilibrium description~\cite{Bamba:2009id}. 
On the other hand, 
in the context of modified gravity theories including $f(R)$ gravity 
a horizon entropy $\hat{S}$ associated with a Noether 
charge has been proposed by Wald~\cite{Wald entropy}. 
The Wald entropy $\hat{S}$ is a local quantity defined 
in terms of quantities on the bifurcate Killing horizon. 
More specifically, it depends on the variation of the Lagrangian 
density of gravitational theories with respect to the Riemann tensor. 
This is equivalent to 
$\hat{S}=A/\left(4G_{\rm eff}\right)$, where $G_{\mathrm{eff}}=G/F$
is the effective gravitational coupling 
in $f(R)$ gravity~\cite{Cognola:2005de}. 
Therefore, we use the Wald entropy in the exponential gravity 
in Eq.~(\ref{eq:2.12}) 
\begin{equation}
\hat{S}=\frac{\left(1-\beta e^{-R/R_{\mathrm{s}}}\right)A}{4G}\,. 
\label{eq:4.7}
\end{equation}
%
In what follows, a hat denotes the quantity in the non-equilibrium 
description of thermodynamics. 

It can be shown that the horizon entropy $S$ in the equilibrium 
description has the following relation with $\hat{S}$ in the 
non-equilibrium description~\cite{Bamba:2009id}: 
\begin{equation}
dS=\frac{1}{1-\beta e^{-R/R_{\mathrm{s}}}} d\hat{S} 
+ \frac{1}{1-\beta e^{-R/R_{\mathrm{s}}}} 
\frac{2H^2+\dot{H}}{4H^2+\dot{H}}\,d_i \hat{S}\,,
\label{eq:4.19}
\end{equation}
with 
\begin{equation}
d_i \hat{S}=-\frac{6\pi}{G} \frac{4H^2+\dot{H}}{H^2} 
\frac{\beta}{R_{\mathrm{s}}} e^{-R/R_{\mathrm{s}}} 
\frac{dR}{R}\,,
\label{eq:4.20}
\end{equation}
where $d_i \hat{S}$ is the new term which can be interpreted as a term 
of entropy produced in the non-equilibrium thermodynamics. 
The difference between $S$ and $\hat{S}$ appears in $f(R)$ gravity due to 
$dF \neq 0$. 
Note that $S$ is identical to $\hat{S}$ in general relativity due to $F=1$. 
{}From Eq.~(\ref{eq:4.19}), we see that the change of the horizon entropy $S$ 
in the equilibrium framework involves the information of 
both $d\hat{S}$ and $d_i \hat{S}$ in the non-equilibrium framework.

In Figs.~7, 8 and 9, 
we show the cosmological evolution of the horizon entropy 
$\hat{S}$ in Eq.~(\ref{eq:4.7}) 
in the non-equilibrium description of thermodynamics 
and $S$ in Eq.~(\ref{eq:4.B-1}) in the equilibrium description 
of thermodynamics for $\beta=1.1$, $1.8$ and $2.5$, respectively. 
In these figures, we illustrate the normalized quantities 
$\bar{\hat{S}} \equiv \hat{S}/S_0$ 
and $\bar{S} \equiv S/S_0$ with 
$S_0 = \pi/\left(GH_0^2\right)$ being 
the present value of the horizon entropy $S$. 
Furthermore, we also depict the evolution of $\bar{H} \equiv H/H_0$. 
We note that as $S \propto H^{-2}$, $S$ increases 
with time as long as $H$ continues to decreases to the de Sitter 
point, in which $H$ becomes a constant. 

In the high $z$ regime ($z \gtrsim 1$), since the deviation of 
the exponential gravity from the $\Lambda\mathrm{CDM}$ model, i.e., 
general relativity, is very small, 
the evolution of $S$ is similar to that of $\hat{S}$. 
In other words, for the high $z$ regime (the higher curvature regime) 
$F(R) = 1-\beta e^{-R/R_{\mathrm{s}}} \approx 1$ 
because $R/R_{\mathrm{s}} \gg 1$. 
As $z$ decreases (and $R$ also decreases), 
the deviation of the exponential gravity from the $\Lambda\mathrm{CDM}$ model 
emerges, i.e., $F(R) <1$ and $F(R)$ decreases. 
Hence, there appears a difference between 
the evolution of $S$ and that of $\hat{S}$. 
Note that $S>\hat{S} \propto F(R)$. 
The present values of $\bar{\hat{S}}$ 
are $\bar{\hat{S}} (z=0) = 0.90$, $0.96$ and $0.99$ for 
$\beta=1.1$, $1.8$ and $2.5$, respectively. 
It is clear from Figs.~7, 8 and 9 that 
both $S$ and $\hat{S}$ globally increases with time for any values of $\beta$. 
This confirms that the second law of thermodynamics on the apparent horizon 
always holds. 
The similar behaviors for both $S$ and $\hat{S}$ have been obtained in 
the Starobinsky's model~\cite{Bamba:2009id}. 
Furthermore, we see that the larger $\beta$ is, 
the closer the evolution of $\bar{\hat{S}}$ is to that of $\bar{S}$.

\begin{figure}[tbp]
\begin{center}
\resizebox{!}{13cm}{
   \includegraphics{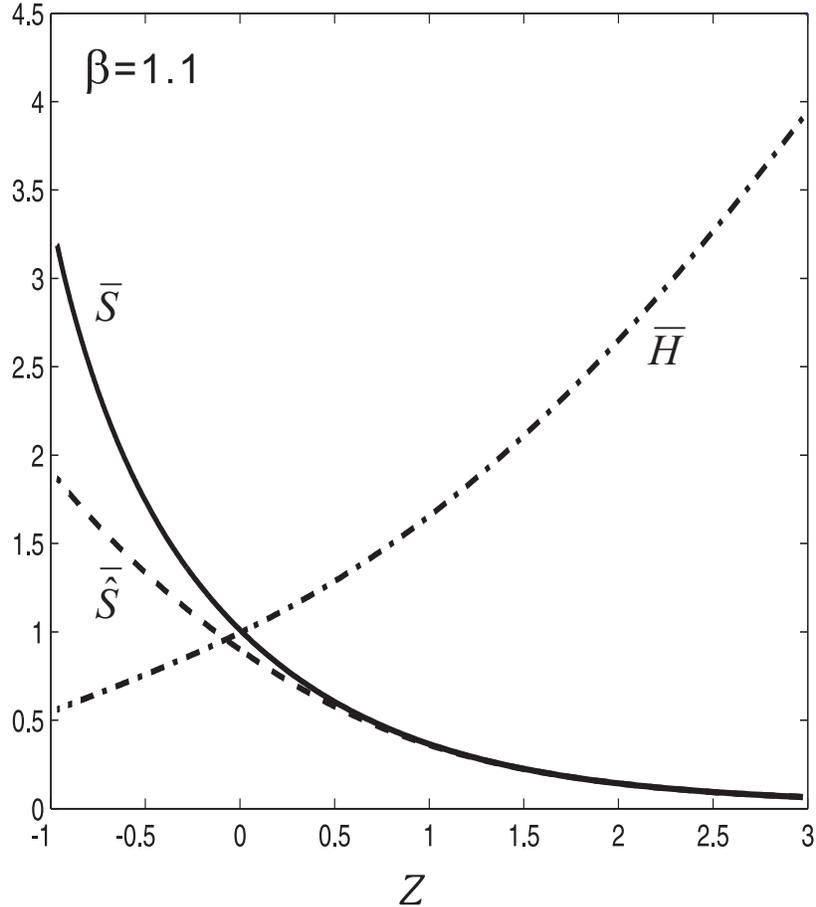}
                  }
\caption{Cosmological evolutions of 
$\bar{S} = S/S_0$ (solid line), $\bar{\hat{S}} = \hat{S}/S_0$ 
(dashed line) and $\bar{H} = H/H_0$ (dashed and single-dotted line) 
as functions of the redshift $z$ 
for $\beta=1.1$. 
}
\end{center}
\label{fg:7}
\end{figure}

\begin{figure}[tbp]
\begin{center}
\resizebox{!}{13cm}{
   \includegraphics{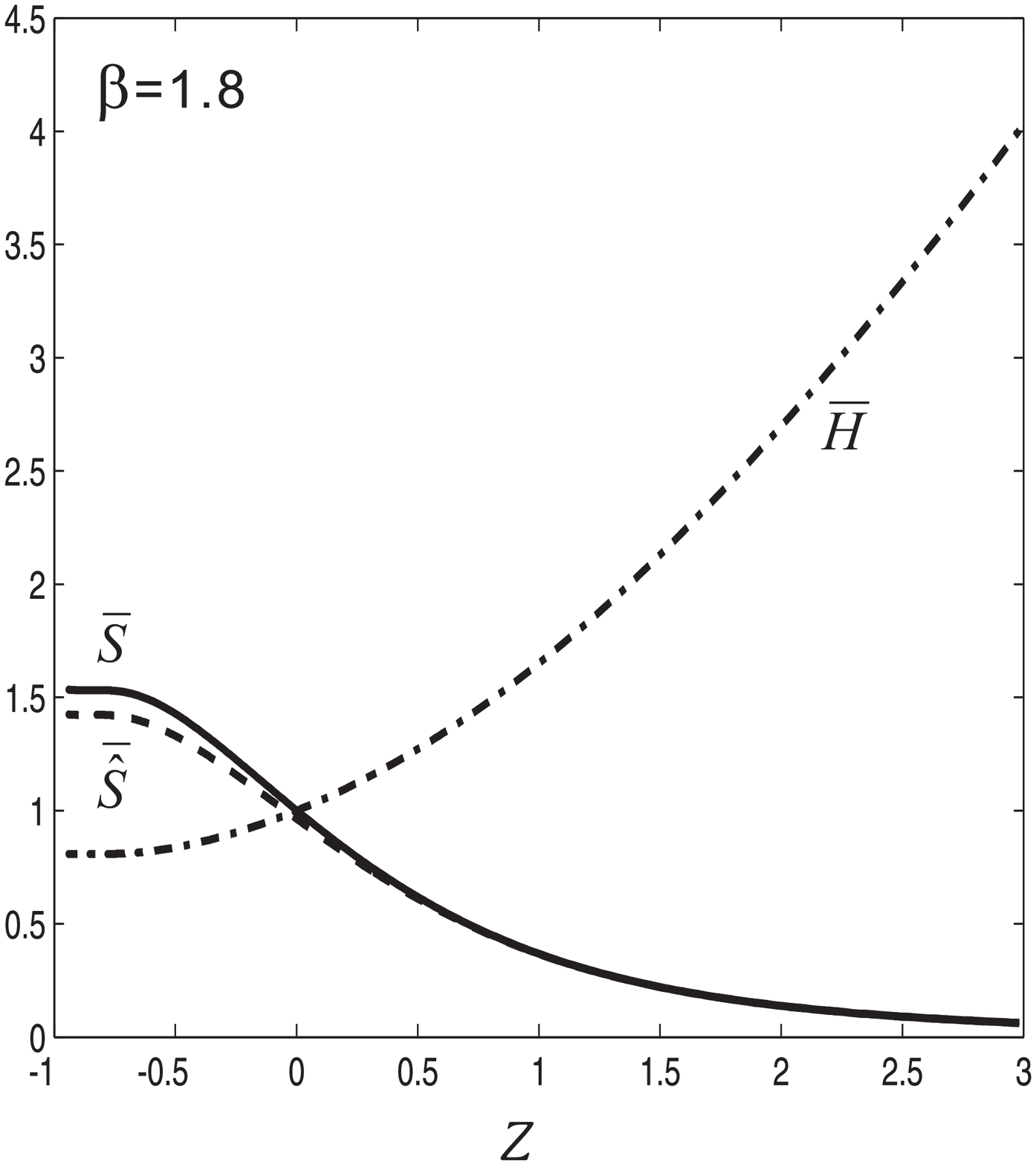}
                  }
\caption{Legend is the same as Fig.~7 but for $\beta=1.8$. 
}
\end{center}
\label{fg:8}
\end{figure}

\begin{figure}[tbp]
\begin{center}
\resizebox{!}{13cm}{
   \includegraphics{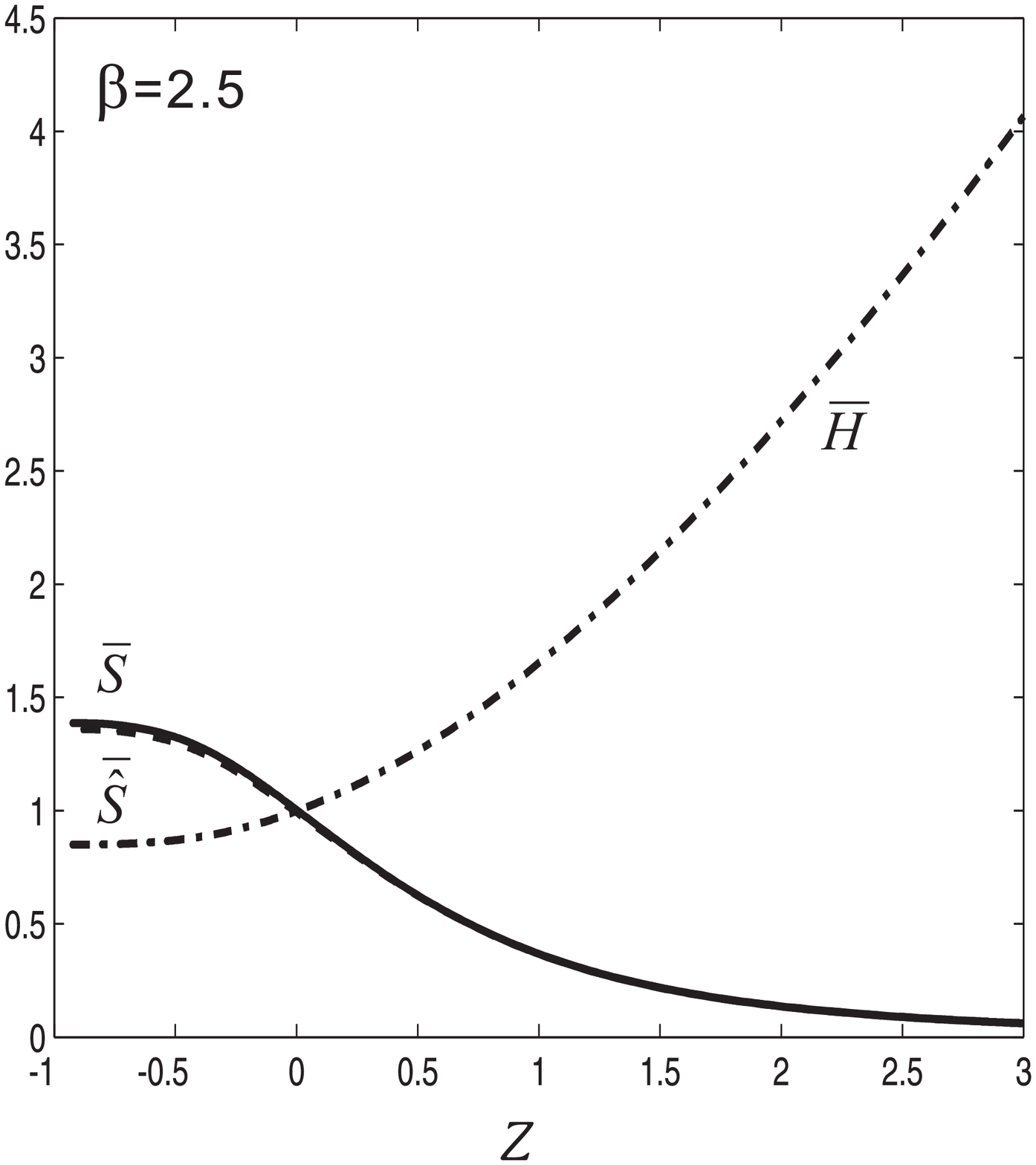}
                  }
\caption{Legend is the same as Fig.~7 but for $\beta=2.5$. 
}
\end{center}
\label{fg:9}
\end{figure}

\section{Conclusions}

In the present paper, we have studied the cosmological evolution in the 
exponential gravity. We have summarized various viability conditions 
and explicitly illustrated that the late-time cosmic 
acceleration can be realized after the matter-dominated stage. 
We have also shown that the crossing of the phantom divide from the phantom 
phase to the non-phantom one can occur and the cosmological horizon entropy 
globally increases with time. 
Phenomenologically, at least in the light of the background cosmological 
evolution, the exponential gravity can be regarded as one of the most 
promising viable modified gravitational theories because 
(a) it satisfies all conditions for the viability; 
(b) in substance it has only one model parameter; and 
(c) both the current cosmic acceleration following the matter-dominated stage 
and the crossing of the phantom divide can be realized. 
%

\section*{Acknowledgments}

We thank Mr. Hayato Motohashi for his important suggestions. 
We are grateful to Professor Shinji Tsujikawa and Dr. Antonio De Felice for 
helpful discussions. 
The work is supported in part by 
the National Science Council of R.O.C. under: 
Grant \#s: NSC-95-2112-M-007-059-MY3 and
NSC-98-2112-M-007-008-MY3
and 
National Tsing Hua University under the Boost Program and Grant \#: 
97N2309F1.


\end{document}